\documentclass[epj]{webofc}
\usepackage[varg]{txfonts}   
\usepackage{subfigure}
\woctitle{MESON2018 - the 15$^\textrm{th}$ International Workshop on Meson Physics}
\begin{document}
\selectlanguage{english}
\title{The pole structure of low energy $\pi N$ scattering amplitudes}

\author{Yu-Fei Wang\inst{1}\fnsep\thanks{\email{1401110076@pku.edu.cn}}
}

\institute{Department of Physics and State Key Laboratory of Nuclear Physics and Technology, \\
Peking University, Beijing, China}
\abstract{This report presents an investigation of the pion-nucleon elastic scattering in low energy region using a production representation of the partial wave $S$ matrix. The phase shifts are separated into contributions from poles and branch cuts, where the left-hand cut term can be evaluated by tree-level covariant baryon chiral perturbation theory. A comparison between the sum of known contributions and the data in $S$- and $P$- wave channels is made. It is found that the known components in $S_{11}$ and $P_{11}$ channels are far from enough to saturate the corresponding experimental data, indicating the existence of low-lying hidden poles. The positions of those hidden poles are figured out and the physics behind them are explored. }
\maketitle
As one of the landmark processes in hadron and nuclear physics, the pion-nucleon ($\pi N$) scattering has been studied for decades. On the one hand, various phenomena are observed by experiments, and a large amount of data are accumulated, see e.g. Refs.~\cite{Arndt:2006,SAID}. On the other hand, there are still many open questions to be studied theoretically, such as the pion-nucleon $\sigma$ term, and the intermediate resonances. Physics in $S_{11}$ and $P_{11}$ (in $L_{2I\ 2J}$ convention) channels may be of particular interest, since the two states $N^*(1535)$ and $N^*(1440)$ cause a lot of puzzles, arousing many theoretical works~\cite{Kaiser:1995cy,Nieves:2001wt,Krehl:1999km,Arndt:1985vj,Ceci:2011ae,Mai:2009ce,Bruns:2010sv}. To gain a clear idea of the physics behind, one demands a method that is model-independent and can work well in low energy region.

Peking University (PKU) representation~\cite{Xiao:2000kx,Zheng:2003rw,Zhou:2004ms,Zhou:2006wm} is such a method which separates the partial wave $S$ matrix of two-body elastic scatterings into different contributions either from poles or branch cuts:
\begin{equation}\label{PKU}
S(s)=\prod_i S_i^p(s)\times S^{cut}(s)\ \mbox{, }
\end{equation}
for $S_i^p$ being the factors corresponding to poles on the first and second Riemann sheets and $s$ being the Mandelstam variable. The $S^{cut}$ in Eq.~\eqref{PKU} is called background term, carrying the information of the left-hand cuts (\textit{l.h.c.}s) and right-hand inelastic cut (\textit{r.h.i.c.}). For the detailed expression of each term, see Ref.~\cite{Zhou:2004ms}. PKU representation is derived from first principles of $S$-matrix theory, thus it is rigorous and model independent for two-body elastic scatterings. Besides, the $S_i^p$ terms are quite sensitive to pole locations that are not too far away from the threshold, while the background term $S^{cut}$ can be evaluated from theories of low energy dynamics, thus PKU representation works well at low energies. In fact, phase shifts corresponding to each term in Eq.~\eqref{PKU} has definite sign: bound states are always negative while poles on the second sheet (virtual states and resonances) are positive; the background term empirically gives negative phase shift (proved in quantum mechanical scattering theory under some assumptions in Ref.~\cite{Regge:1958ft}). Due to those advantages, one can utilize PKU representation to dig out hidden contributions. The verification of the existence of $\sigma$ ($f_0(500)$) particle in Ref.~\cite{Xiao:2000kx} is a vivid example: because of the negative definite background contribution, the $\sigma$ pole undoubtedly exists to give a sizeable positive contribution to match the data. Moreover, comparisons between PKU representation and some conventional unitarization approaches can be found in Ref.~\cite{Qin:2002hk}.

In what follows PKU representation is applied to $\pi N$ scatterings to make a comparison between known contributions and the phase shift data. The known contributions mainly come from the resonances observed by experiments (taken from the results in Ref.~\cite{Anisovich:2011fc}) and the \textit{l.h.c.}s; \textit{r.h.i.c.} is omitted temporarily since its contribution is small in the low energy region. However resonances in experiments are often third sheet poles, since only the total width can be read out from the line shape. Here narrow width approximation is adopted to obtain the corresponding shadow poles on second sheet, i.e. $\sqrt{s}^{II}=M_r-\frac{\Gamma_1-\Gamma_2}{2}i$, where $M_r$ is the mass of the resonance and $\Gamma_1,\ \Gamma_2$ label the partial decay width to $\pi N$ channel and inelastic channels respectively. As for the \textit{l.h.c.}s, covariant baryon chiral perturbation theory (BChPT) is employed. Recently the perturbative calculations are carried out up to $\mathcal{O}(p^4)$ level~\cite{Alarcon:2011kh,AlarconA,Chen:2012nx,Yao:2016vbz,Siemens:2016jwj}, however for simplicity the result up to $\mathcal{O}(p^2)$ level is used in this report. The Lagrangians up to $\mathcal{O}(p^2)$ can be found in Ref.~\cite{Fettes:2000gb}. The background term in Eq.~\eqref{PKU} can be written as $S^{cut}=e^{2i\rho(s)f(s)}$, with
\begin{equation}\label{fsint}
f(s)=-\frac{s}{\pi}\int_{s_{c}}^{(M-m)^2} \frac{\ln|S(w)|dw}{2\rho(w)w(w-s)}+\frac{s}{\pi}\int_{(M^2-m^2)^2/M^2}^{2m^2+M^2} \frac{\text{Arg}[S(w)]dw}{2iw\rho(w)(w-s)}\ \mbox{, }
\end{equation}
where $m$ and $M$ stand for the masses of the pions and the nucleons respectively, and the kinematic factor is given by $\rho(s)=\frac{\sqrt{s-(M+m)^2}\sqrt{s-(M-m)^2}}{s}$. The $S$ matrix in Eq.~\eqref{fsint} is $S=1+2i\rho T$ and $T$ is the partial wave perturbative amplitude obtained by BChPT. Actually the second term in Eq.~\eqref{fsint} is originated from $u$ channel nucleon exchange which is numerically very small, and the first term corresponds to the kinematic left-hand cut $(-\infty,(M-m)^2]$, which contributes to the phase shift negatively. Due to the fact that BChPT can not work at high energy region, a cut-off parameter $s_c$ is assigned to the integral.

\begin{figure}[h]
\centering
\subfigure[]{
\label{p2PKU:subfig:S11}
\scalebox{0.8}[0.8]{\includegraphics[width=0.6\textwidth]{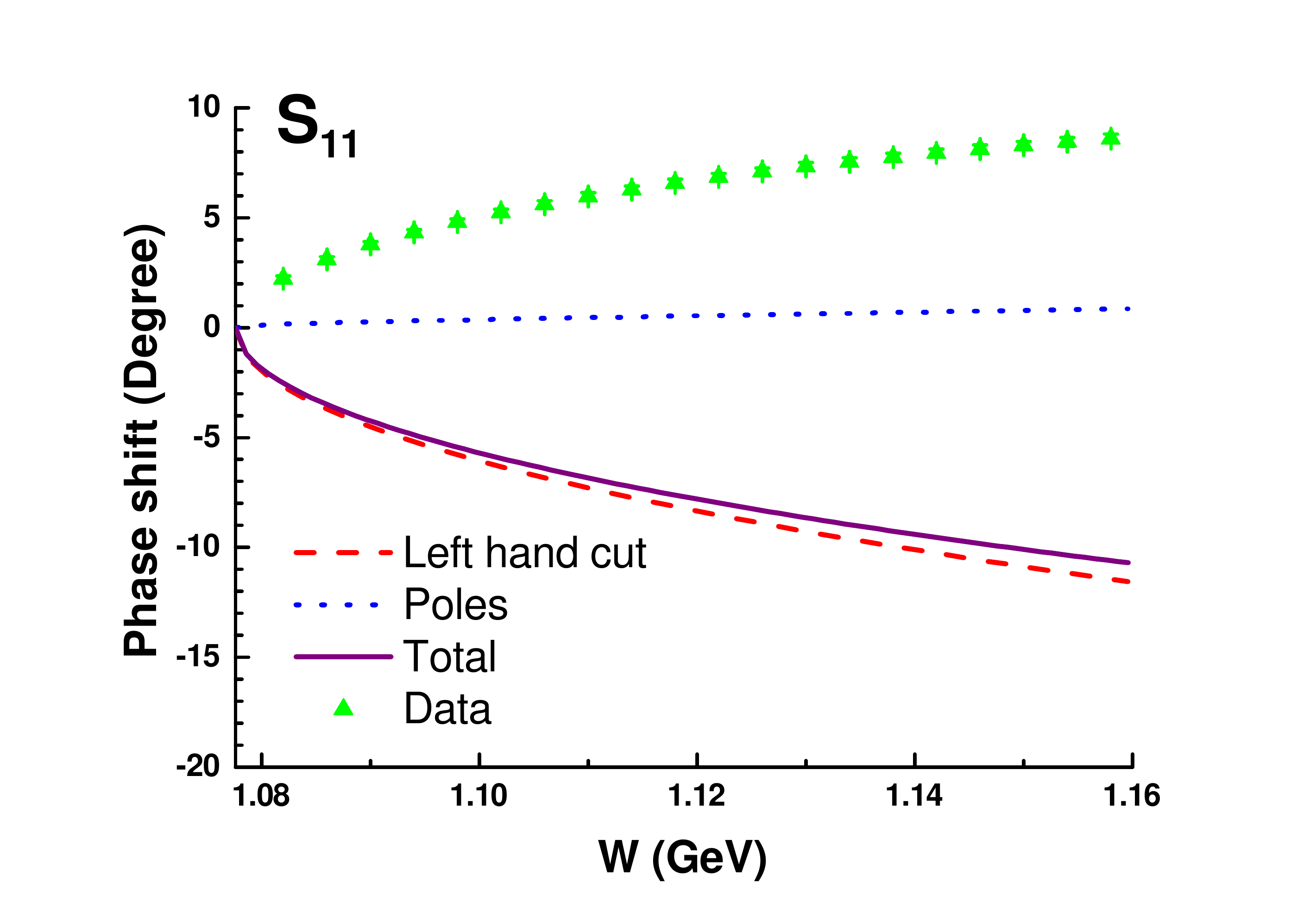}}}
\subfigure[]{
\label{p2PKU:subfig:S31}
\scalebox{0.8}[0.8]{\includegraphics[width=0.6\textwidth]{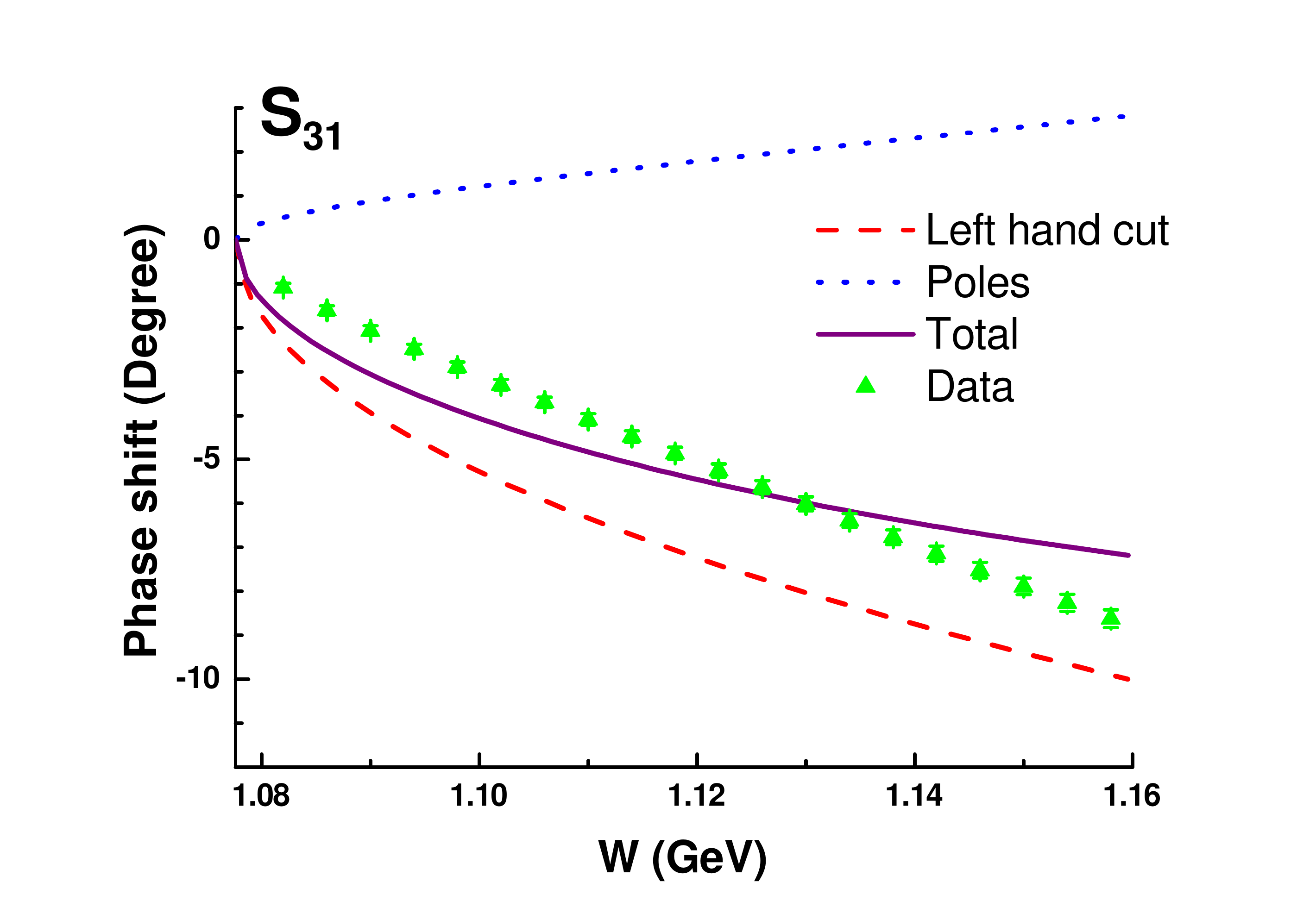}}}
\subfigure[]{
\label{p2PKU:subfig:P11}
\scalebox{0.8}[0.8]{\includegraphics[width=0.6\textwidth]{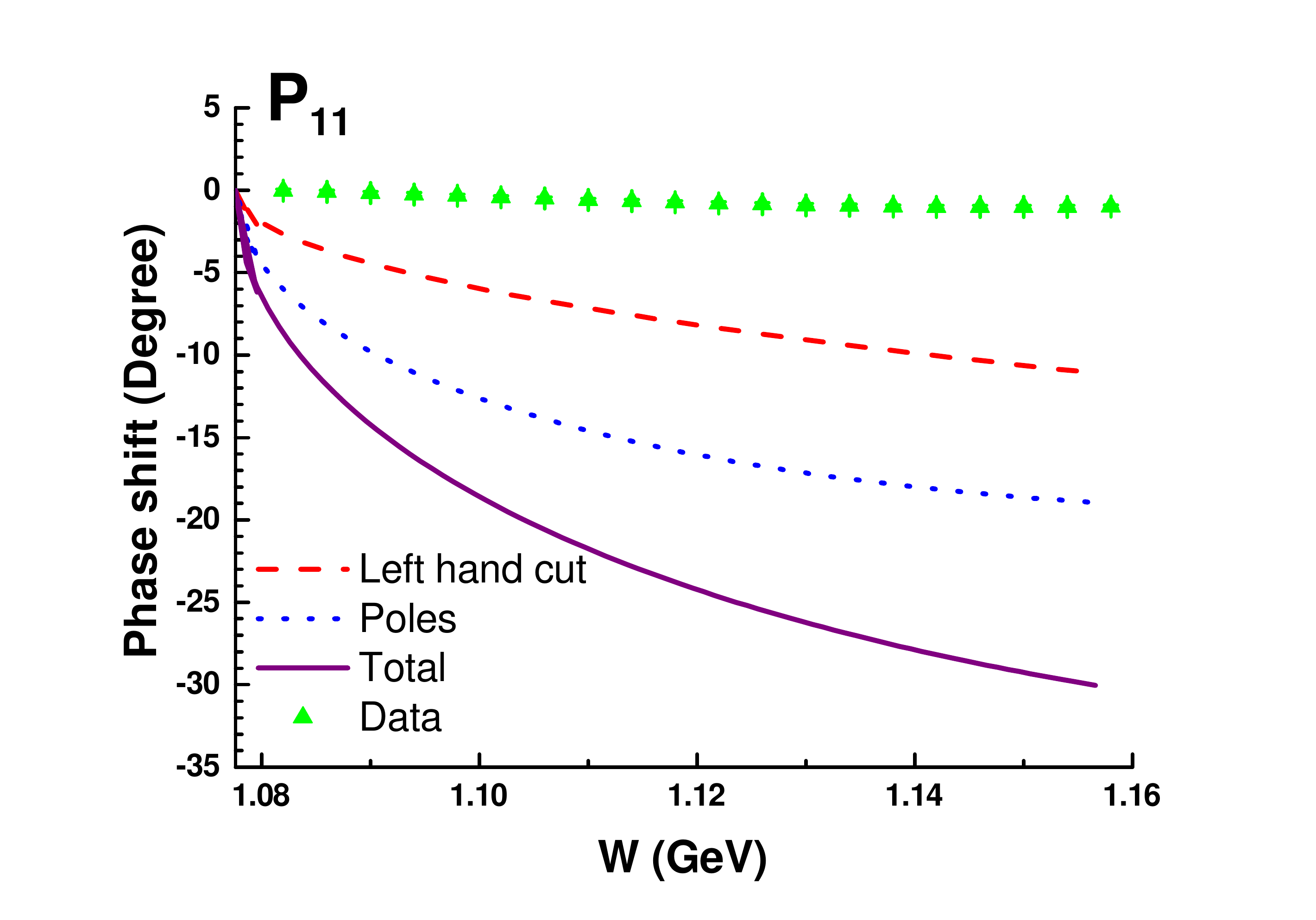}}}
\subfigure[]{
\label{p2PKU:subfig:P31}
\scalebox{0.8}[0.8]{\includegraphics[width=0.6\textwidth]{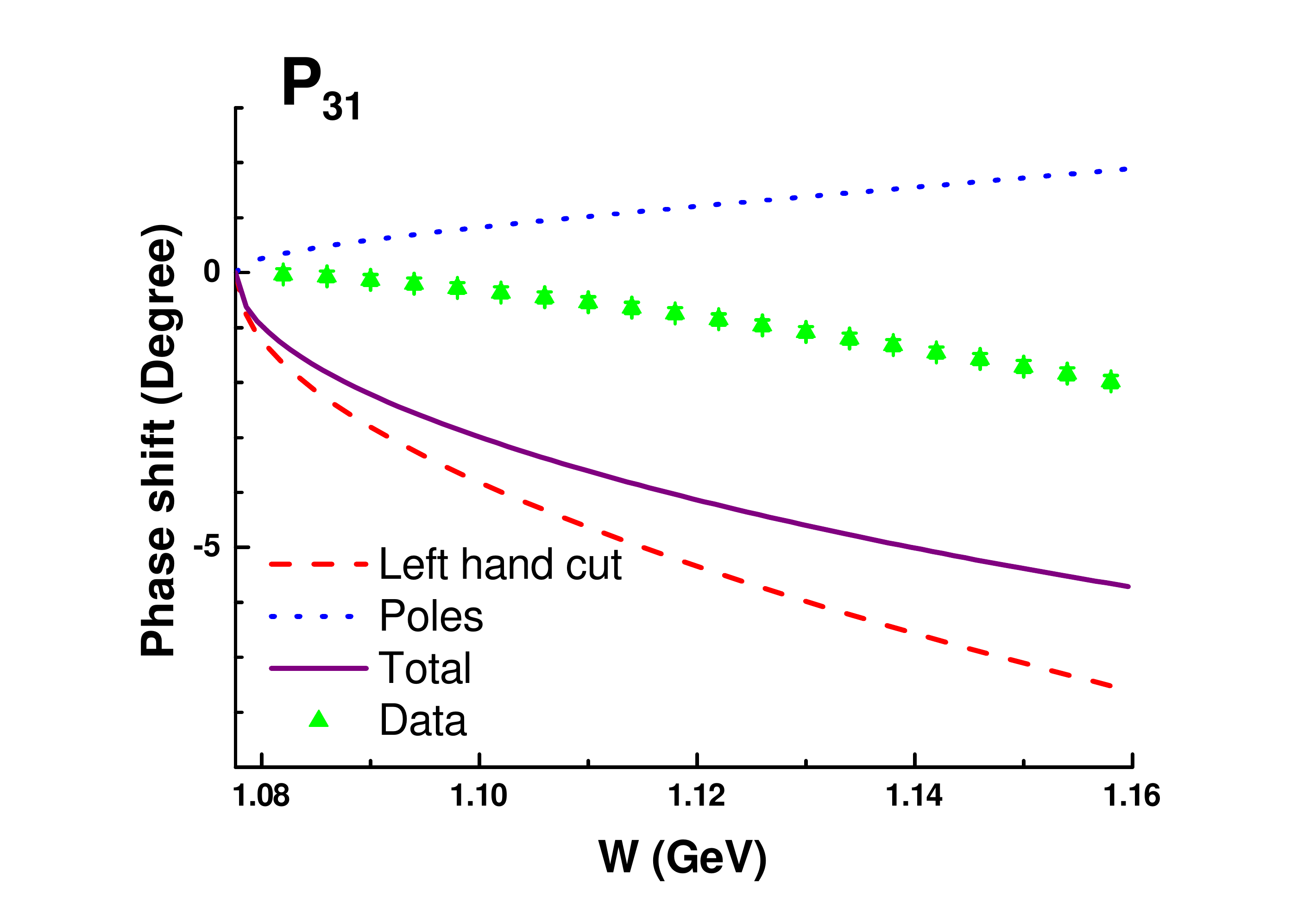}}}
\subfigure[]{
\label{p2PKU:subfig:P13}
\scalebox{0.8}[0.8]{\includegraphics[width=0.6\textwidth]{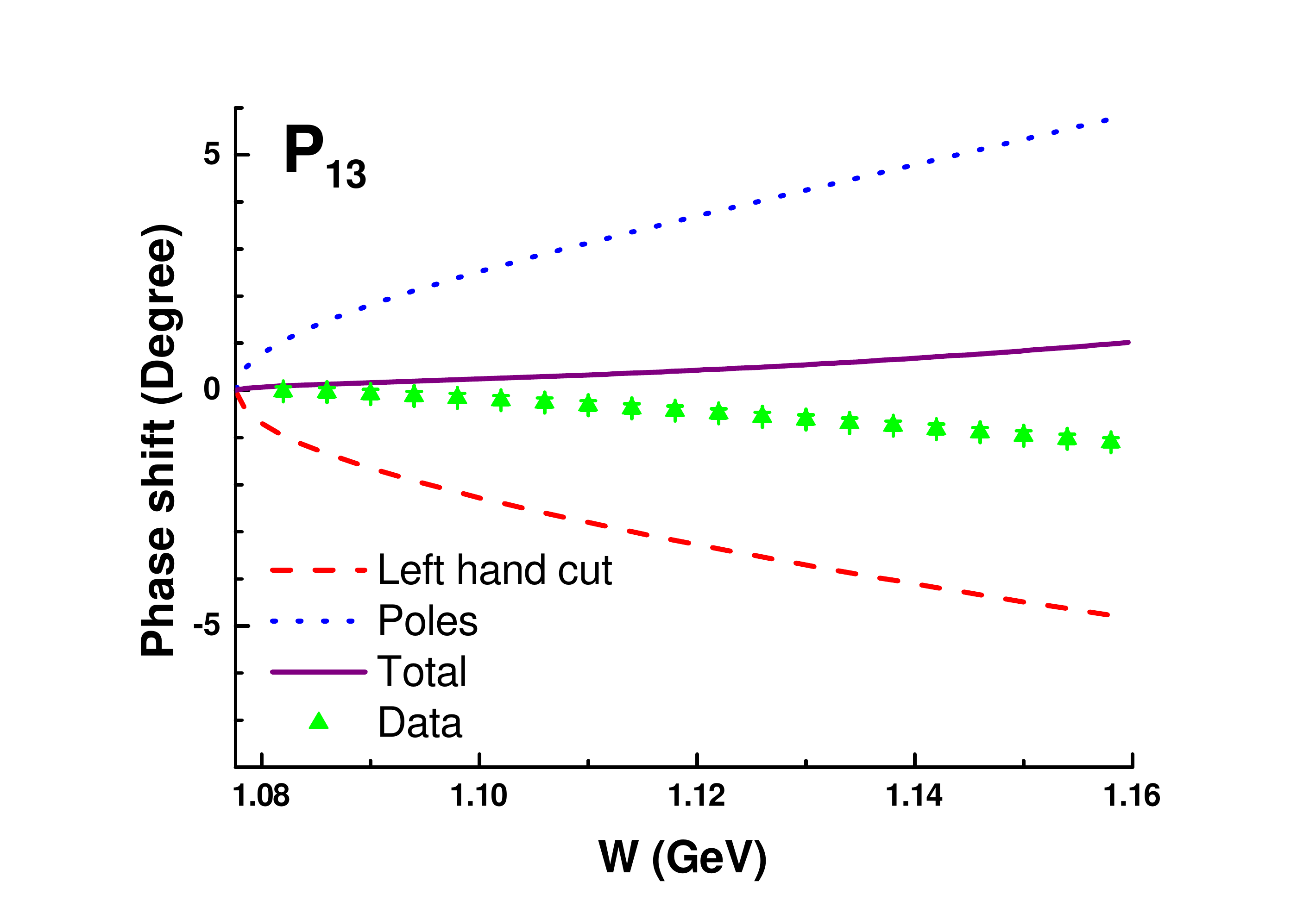}}}
\subfigure[]{
\label{p2PKU:subfig:P33}
\scalebox{0.8}[0.8]{\includegraphics[width=0.6\textwidth]{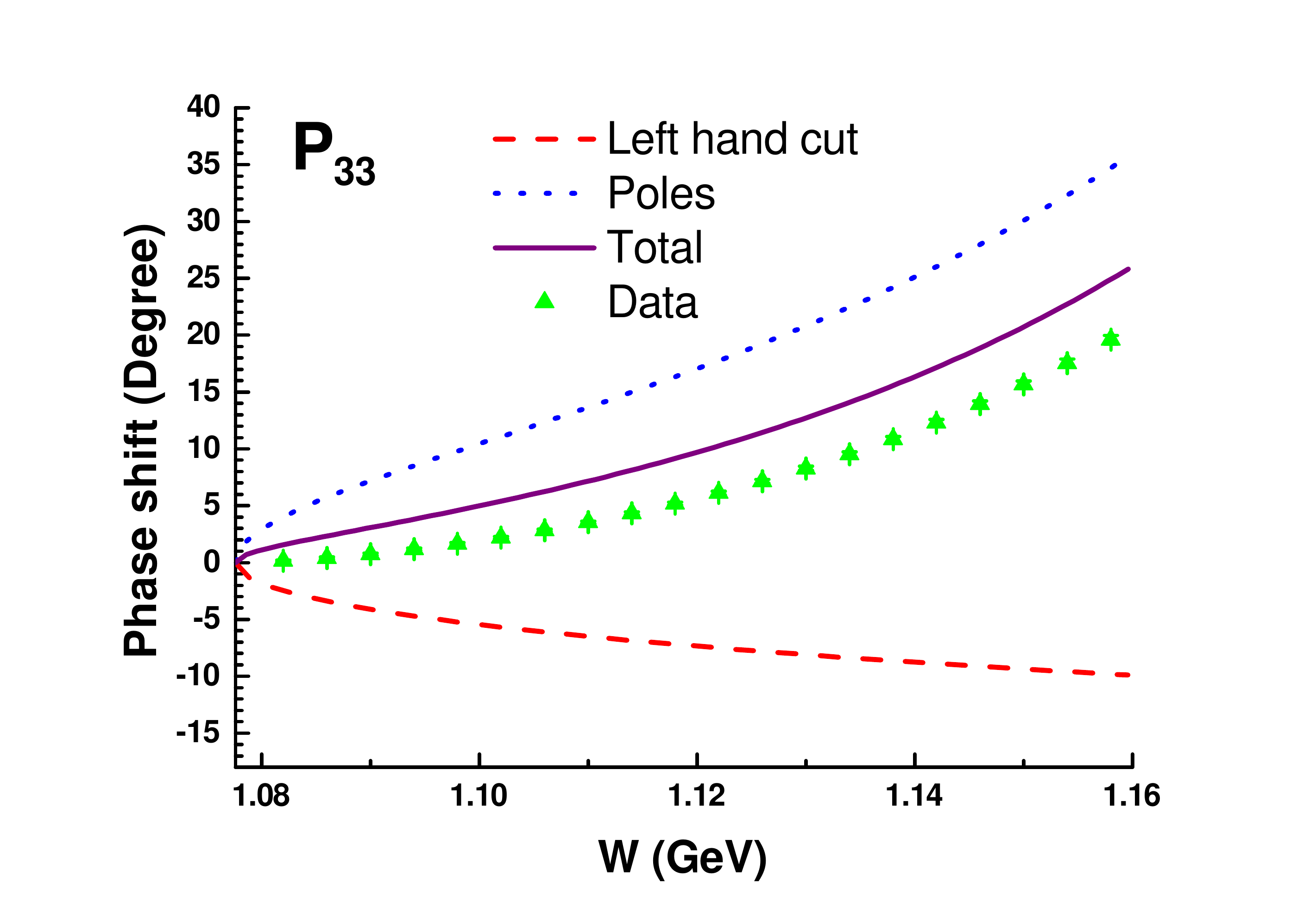}}}
\caption{Tree level PKU representation analyses of the $\pi N$ elastic scattering in $S$- and $P$- waves. }
\label{p2PKU}
\end{figure}

\begin{figure}
\centering
\includegraphics[width=6cm,clip]{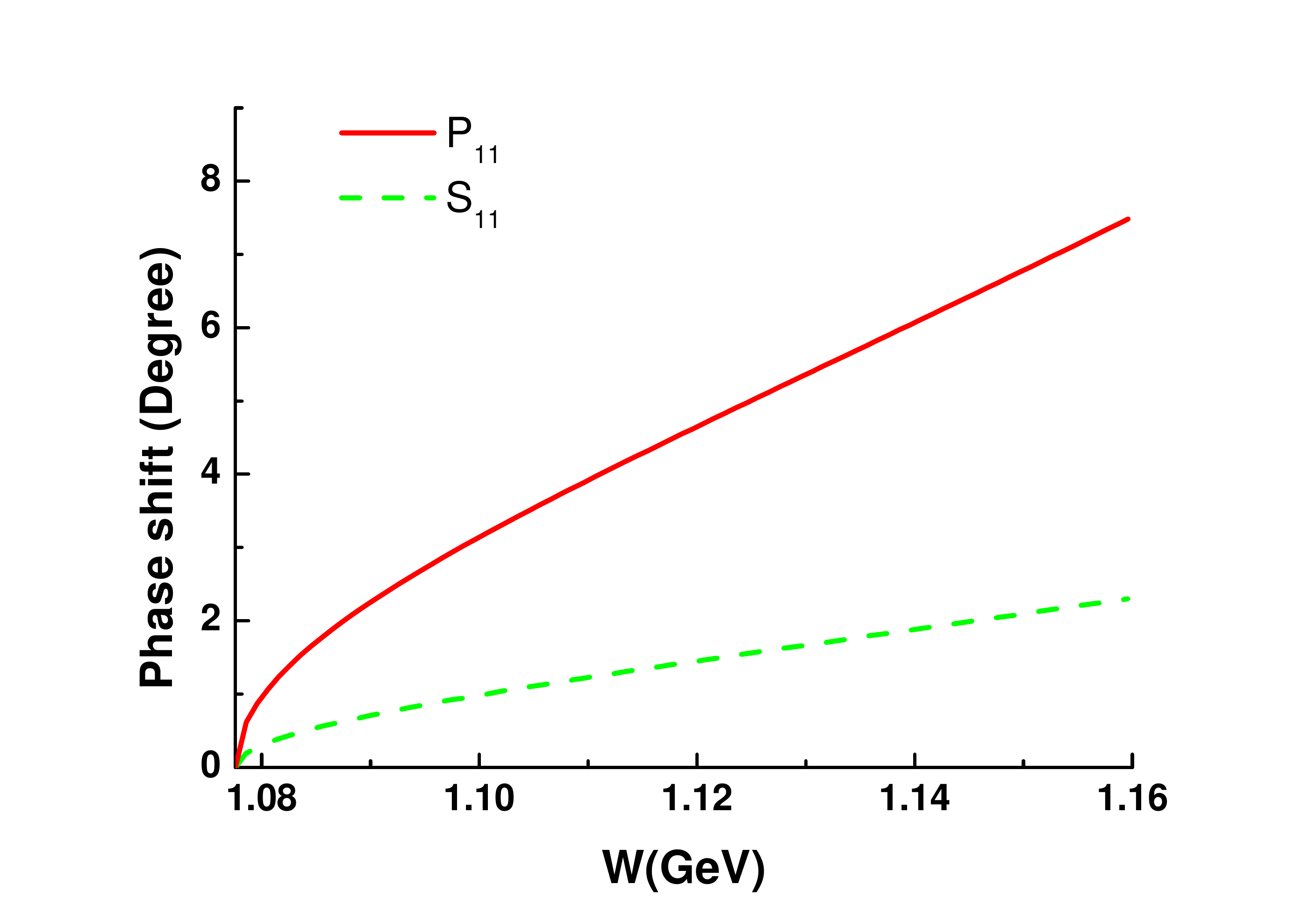}
\caption{The phase shifts from \textit{r.h.i.c.} in $S_{11}$ and $P_{11}$ channels. }
\label{S11P11rhc}
\end{figure}

To proceed, the masses and $\mathcal{O}(p^1)$ constants are set to the values taken from Ref.~\cite{Chen:2012nx}, while $\mathcal{O}(p^2)$ coupling constants are determined by a $K$ matrix fit to the data in Ref.~\cite{SAID} (with errors assigned using the method following Ref.~\cite{Alarcon:2011kh}). The cut-off parameter is set to $s_c=-0.08$ GeV$^2$ according to the $N^*(1440)$ shadow pole position. Thus the comparison between known contributions and the data in six $S$- and $P$- wave channels are shown in Fig.~\ref{p2PKU}.
Firstly one can see that some of the known poles have significant contributions to the phase shifts, e.g. the large and positive phase shift in $P_{33}$ channel mainly comes from the $\Delta(1232)$ pole, and in $P_{11}$ channel the nucleon itself is a bound state of pion and nucleon, so it gives a large and negative contribution. However, significant disagreements can be seen in $S_{11}$ and $P_{11}$ channels -- apart from the known contributions, some important positive components have been missed, similar to what happens in $I=0,J=0$ channel of the $\pi\pi$ scattering, as revealed in Ref.~\cite{Xiao:2000kx}. Note that disagreements also exist in other channels, but they are minor and quantitative. On the contrary, discrepancies in $S_{11}$ and $P_{11}$ channels are crucial -- even if one switches off the \textit{l.h.c.}s, they remains there. Hence one can not erase the disagreements by changing the details related to the \textit{l.h.c.}s, e.g. values of the constants, chiral order etc. Furthermore, \textit{r.h.i.c.} can be evaluated using the data of inelasticity $\eta$ given in Ref.~\cite{SAID}. As shown in Fig.~\ref{S11P11rhc}, the \textit{r.h.i.c.} contributions are numerically important but still far from enough to match the data.

Therefore, extra poles are added to the $S$ matrix in PKU representation and their locations are determined by fitting to the data. In $P_{11}$ channel the $P$- wave threshold constraint (the phase shift $\delta\sim\mathcal{O}(k^3)$ for $3$- momentum $k$) is also taken into consideration. The fit results in a near-threshold virtual state, though good fit quality requires a larger $s_c$ value. For example, when $s_c=-9.00$ GeV$^2$ the virtual state locates at $980$ MeV. Such a virtual state lying above the nucleon pole is actually needed physically, since it can be regarded as the kinematic companionate pole of the nucleon. On the first Riemann sheet, $S(s)$ can be expanded in the vicinity of nucleon pole $S(s)\sim \frac{r_0}{s-M^2}+b_0+\mathcal{O}(s-M^2)\ \mbox{, }$any nonzero $b_0$ can give rise to a zero of $S(s)$, which corresponds to a virtual pole on the second sheet.

As for $S_{11}$ channel, the fit results in a ``crazy resonance'', i.e. a resonance below threshold. Its locations with respect to different $s_c$ values are listed in Table.~\ref{S11pole}.
\begin{table}[!htb]
\centering
\caption{The $S_{11}$ hidden pole fit with different choices of $s_{c}$. }
\label{S11pole}       
\begin{tabular}{ll}
\hline
$s_{c}$ (GeV$^2$)  & Pole position (GeV) \\
\hline
$-0.08$ & $0.808-0.055i$\\
$-1.00$ & $0.822-0.139i$\\
$-9.00$ &$0.883-0.195i$\\
$\infty$ &$0.914-0.205i$\\
\hline
\end{tabular}
\end{table}
That resonance is likely to be a potential-nature state, since in $S_{11}$ channel the nucleon do not appear as a $s$ channel intermediate particle. One can use the square-well potential as a toy model to figure out the phase shift by solving Schr\"{o}dinger equation. It is found that when the square-well range $L=0.829$ fm and the depth $V_0=144$ MeV, the data can be well described, and a pole at $0.872-0.316i$ GeV can be found in the mean time, which is close to the fit results of PKU representation.

To summarize, in this report the production representation of the partial wave $S$ matrix, i.e. PKU representation, is applied to $\pi N$ elastic scatterings. It is found that in $S_{11}$ and $P_{11}$ channels the known contributions, i.e. the branch cuts and the poles already observed, are far from enough to match the data. Through further investigation, two extra hidden states are found: a virtual state induced by the nucleon pole in $P_{11}$ channel, and a resonance lying below $\pi N$ threshold in $S_{11}$ channel that is likely to be a state generated by potential mechanism. The existence of such two states are not sensitive to the details of the calculation.

Follow-up of this work is underway. The $\mathcal{O}(p^3)$ calculation of the left-hand cuts shows no qualitative difference compared with $\mathcal{O}(p^2)$ results, while higher chiral order improves the fit qualities of each channel.

\begin{acknowledgement}
I would like to thank Han-Qing Zheng and De-Liang Yao for helpful advices. This work is supported in part by National Nature Science Foundations of China (NSFC) under Contracts No. 10925522, No. 11021092, No. 11575052 and No. 11105038.
\end{acknowledgement}

\end{document}